\documentclass[twocolumn,showpacs,floatfix, aps, nofootinbib]{revtex4}
\usepackage{amssymb,amsmath}
\usepackage[dvips]{graphics,color}
\usepackage{epsfig}\usepackage{float}
\newcommand{\ba}{\begin{eqnarray}}
\newcommand{\ea}{\end{eqnarray}}
\newcommand{\bege}{\begin{equation}}
\newcommand{\bpartial}{\mathop{\partial\kern -4pt\raisebox{.8pt}{$|$}}}
\newcommand{\enge}{\end{equation}}
\newcommand{\beq}{\begin{eqnarray}}
\newcommand{\benu}{\begin{enumerate}}

\newcommand{\enu}{\end{enumerate}}
\newcommand{\eeq}{\end{eqnarray}}

\newcommand{\lie}{{\bf \pounds}_{n}}

\begin{document}

\title{Gravitational Waves in Braneworld Scenarios with AdS Background}
\author{A. C. Amaro de Faria}
\email{carlos.amaro@ufabc.edu.br} \affiliation{ Centro de
Matem\'atica, Computa\c c\~ao e Cogni\c c\~ao, Universidade
Federal do ABC, 09210-170, Santo Andr\'e, SP, Brazil}
\author{R. da Rocha}
\email{roldao.rocha@ufabc.edu.br} \affiliation{ Centro de
Matem\'atica, Computa\c c\~ao e Cogni\c c\~ao, Universidade
Federal do ABC, 09210-170, Santo Andr\'e, SP, Brazil}
\author{M. E. S. Alves}
\email{alvesmes@unifei.edu.br}
\affiliation{Instituto de Ci\^encias Exatas, Universidade Federal de Itajub\'a, 37500-903 Itajub\'a, MG, Brazil}
\author{J. C. N. de Araujo}
\email{jcarlos@das.inpe.br}
\affiliation{INPE - Instituto Nacional de Pesquisas Espaciais - Divis\~ao
de Astrof\'isica,
\\Av. dos Astronautas 1758, S\~ao Jos\'e dos Campos, 12227-010 SP, Brazil}

\begin{abstract}
In this paper, we investigate gravitational waves as metric
perturbations around a general warped $5$-dimensional background. We find an analytical solution in Randall-Sundrum braneworld model and analyze the implications of
braneworld models in the gravitational waves propagation.
\end{abstract}
\maketitle


\section{Introduction}

Braneworld scenarios wherein the observable universe is trapped on a brane,  embedded in some higher-dimensional spacetime, can explain the hierarchy problem
\cite{ad1,Antoniadis19981,Antoniadis19982,Antoniadis19983,Antoniadis19984,Antoniadis19985,ad3,Shiu1998}. Braneworld models also provide
alternatives to Kaluza-Klein compactification, where the topology has a compactification radius of the Planck
length order. These possibilities come from developments in
non-perturbative string theory, wherein the so-called $D$-branes are elicited and evinced as ($D+1$)-dimensional manifolds in which the standard
model of particles and fields can be consistently confined.  A plausible reason for the weak appearance of the gravitational force, with respect to other forces, can be its dilution in a higher-dimensional
bulk, where $D$-branes \cite{gr,zwi, zwi1, zwi2, Townsend} are embedded.  $D$-branes
are good candidates for braneworlds because among some outstanding features they possess gauge symmetries \cite{zwi, zwi1, zwi2}. The
gauge symmetry arises from open strings, which can collide to form a closed string --- which simplest excitation modes
correspond precisely to gravitons --- that can leak into a higher-dimensional
bulk.
The possibility concerning the existence of extra dimensions may still ascertain
physical aspects on string theory and $D$-branes.
An alternative approach to the extra dimension compactification was pointed by Randall-Sundrum braneworld formalism \cite{RS,RS1}, where the electromagnetic, weak, and strong forces, together with all the matter
in the universe as well are confined on a 3-brane, and only gravitons would be allowed to leave the surface
 and move into the AdS$_5$ bulk.

At high energies,
 significant changes are introduced in gravitational dynamics, forcing general relativity to be emulated and overcame by a quantum gravity theory \cite{rov}.
  Randall-Sundrum braneworld
  models \cite{RS,RS1}  induce a volcano barrier shaped effective potential for gravitons around the brane \cite{Likken}.
The corresponding spectrum of gravitational perturbations has a massless bound state on the brane,
and a continuum of bulk modes with suppressed couplings to brane fields, which introduces small
corrections at short distances.

Although this alternative removes the hierarchy between the weak  and
the Planck scales, there is still a hierarchy between the weak and
the compactification scales. However, the geometries arising from the Horava-Witten theory
\cite{Horava1996, Horava19961} can explain the origin of this
resulting hierarchy, as was treated by Randall-Sundrum
\cite{RS,RS1}.

In this perspective, gravity can propagate in the
higher-dimensional manifold without modifying Newton's inverse
square law \cite{RS,RS1}. This is possible because a
curved background could generate higher-dimensional modes
of the graviton in the extra dimension.

In this context,
the aim of the present work is to analyze
gravitational waves as metric perturbations around a general 5-dimensional metric and also to reveal
their prominent characteristics in braneworld scenarios, obtained in a background geometry. Such geometry associated would arise from a specific model
characterized by a suitable action, which is equivalent to consider a
general perturbation on the curved background. The
gravitational waves propagation depends also on the extra dimension, in this formalism.

The paper is organized as follows: the background geometry on
which the perturbations are expanded is briefly described in
Section \ref{back space}. In Section \ref{GW 5D}, in order to provide analytic solutions for the gravitational waves, we delve into the Randall-Sundrum formalism and find explicitly analytical solutions for the gravitational waves
in this braneworld scenario, forthwith in full compliance to the standard usual 4-dimensional Minkowski solutions, which are the limit of the braneworld gravitational waves analytic solutions when the extra dimension tends to zero.
We summarize our concluding remarks in Section \ref{concl}.

\section{The background spacetime}\label{back space}

Hereon $\{e_\mu\}$, {\footnotesize{$\mu = 0,1,2,3$}} [$\{e_A\}$, {\footnotesize{$A=0,1,2,3,4$}}] denotes a basis for the tangent space $T_xM$ at a point $x$ in a 3-brane
$M$ embedded in a bulk.
Naturally the cotangent space at $x$ has an orthonormal basis $\{\theta^\mu\}$ [$\{\theta^A\}$] such that $\theta^\mu(e_\nu) = \delta^\mu_{\;\,\nu}$.
If we choose a local coordinate chart, it is possible to represent $e_A = \partial/\partial x^A$ and $\theta^A = dx^A$.

Take $n = n^Ae_A$ a  vector orthogonal to $T_xM$ and let $y$ be the Gaussian coordinate orthogonal to $T_xM$, indicating
how an observer upheavals out the brane into the bulk. In particular, $n_Adx^A = dy$.
A vector field $v = x^Ae_A$ in the  bulk  is split into components in the brane and orthogonal to the brane, respectively as
 $v = x^\mu e_\mu + ye_4 = (x^{\mu},y)$. Since the bulk is endowed with a
metric ${}^{(5)}g = {}^{(5)}g_{AB}dx^A dx^B$,  the components of the metric in the brane and in the bulk, denoted respectively by  $g_{AB}$ and
 $^{(5)}g_{AB}$, are related by
\begin{equation}\label{neo}
{}^{(5)}g_{AB} = g_{AB} + n_An_B.
\end{equation}
The extrinsic curvature components can be defined via the
Lie derivative as $K_{AB}=\frac{1}{2}\lie g_{AB}$.
The extrinsic curvature of the brane localized at $y=0$  describes the embedding of the brane in the bulk, and projects the bulk Riemann tensor on the brane as given by $R_{ABCD} = {}^{(5)}R_{EFGH} g_{\;A}^{E}g_{\;B}^{F}g_{\;C}^{G}g_{\;D}^{H} + 2K_{A[C}K_{D]B}$.

A background metric containing a warp factor which is a function of the extra dimension
is now considered:
\begin{equation}\label{back metric}
{}^{(5)}ds^2 = {}^{(5)}g_{AB}dx^Adx^B = e^{2\zeta(y)}\eta_{\mu\nu}dx^\mu  dx^\nu + dy^2,
\end{equation}
The behavior of the so-called warped factor $\zeta(y)$ can be
derived from a general action \cite{Hawking2000}
\begin{equation}\label{general action}
S_{gra} = S_{EH} + S_{GH} + S_1 + S_2 + \cdots
\end{equation}
where the additional terms beyond $S_2$ make the action
to be finite, since it presents logarithmic divergences
\cite{Henning2000,Tseytlin1998,Henning1998}. In the most general case of a $(D+1)$-dimensional spacetime,
the first term is the usual Einstein-Hilbert action:
\begin{equation}\label{EH action}
S_{EH} = -\frac{1}{16\pi G_{D+1}} \int d^{D+1}x\sqrt{g_{D+1}}\left(R + \frac{D(D-1)}{\ell^2}\right),
\end{equation}
where $G_{D+1}$ denotes the ($D+1$)-dimensional Newton constant
\cite{maartens} $G_{D+1} = M^{1-D}_{D+1}G$,
where $M_{D+1}$ denotes the $(D+1)$-dimensional Planck mass, $G$ denotes the 4-dimensional gravitational constant, and $D$ denotes
the number of spatial dimensions.
Also, $R$ denotes the curvature scalar of the boundary, the metric $g_{D+1}$ is the
AdS$_{D+1}$ metric and $\ell$ denotes the AdS$_{D+1}$ curvature radius.

The second term in the action (\ref{general action}) is given by
\begin{equation}
S_{GH} = -\frac{1}{8\pi G_{D+1}} \int d^Dx \;\;\sqrt{h} K,
\end{equation}
where $K$ is related to the boundary curvature and $h$ denotes
the determinant of the induced metric \cite{Gibbons1977}.

The first two counter terms in the action (\ref{general
action}) are given by
\cite{Henning2000,Balasu1999,Emparan1999,Kraus1999}:
\begin{equation}
S_1 = \frac{D-1}{8\pi G_{D+1}\ell}\int d^Dx\;\;\sqrt{h},
\end{equation}
and
\begin{equation}
S_2 = \frac{\ell}{16\pi G_{D+1}(D-2)}\int d^Dx\;\;\sqrt{h}R,
\end{equation}

In the particular case of Randall-Sundrum braneworlds, gravity is localized in the brane by warped compactification, and what precludes gravity from
leaking into the extra dimension $y$ at low energies is a negative
bulk cosmological constant, $\Lambda_5 = -\frac{6}{\ell^2}=-6\mu^2$,
where $\mu$ denotes the
corresponding energy scale. The curvature radius determines the
magnitude of the Riemann tensor \cite{maartens}:
 \begin{equation}
{}^{(5)}R_{ABCD}=-\frac{1}{\ell^2}\left({}^{(5)}g_{AC} {}^{(5)}g_{BD} -
{}^{(5)}g_{AD} {}^{(5)} g_{BC} \right).
 \end{equation}
The bulk cosmological constant $\Lambda_5$ imposes
the gravitational field to be closer to the brane. In Gaussian normal coordinates $X^A=(x^\mu,y)$ the AdS$_5$ metric takes the form
 \begin{equation}
{}^{(5)} ds^2=e^{-2|y|/\ell} \eta_{\mu\nu}dx^\mu dx^\nu + dy^2\,,
 \end{equation}
where $\eta_{\mu\nu}$ denotes the Minkowski metric. The exponential warp
factor reflects the confining role of the bulk cosmological
constant. The $\mathbb{Z}_2$ symmetry about the brane at $y=0$ is
incorporated via the $|y|$ term. In the bulk, this metric is a
solution of the 5-dimensional Einstein equations, $ {}^{(5)}G_{AB} = -\Lambda_5\,\, {}^{(5)} g_{AB}$.
  The brane is a flat
Minkowski spacetime,
with self-gravity in the form of brane tension.

The solution of Eq.(\ref{EH action}) in the case of $D=4$ presents the following warp factor (\ref{back
metric}) for the case of the Randall-Sundrum braneworld model\footnote{For hybrid compactification, otherwise the warp factor is given by $\zeta(y) = -k|y|$.}:
\begin{equation}\label{warp}
\zeta(y) = -ky,
\end{equation}
where $k=\ell^{-1}$ is the inverse of the AdS$_5$ radius $\ell$.

\section{Gravitational waves solution in braneworld models}\label{GW
5D}

Let us consider small perturbations around the background metric $\eta_{\mu\nu} \mapsto \eta_{\mu\nu} + h_{\mu\nu}$, where $h_{\mu\nu} \ll 1$
in Randall-Sundrum braneworld scenario
\begin{equation}
ds^2 = e^{2\zeta(y)}(\eta_{\mu\nu} + h_{\mu\nu})dx^\mu dx^\nu + dy^2,
\end{equation}

In Gaussian coordinates the Lie derivative $\lie$ equals $\partial/\partial y$. Such approach was comprehensively used in different applications  of braneworld formalism, as in \cite{nossos, nossoss1, nossoss2, maartens, nossos1, nossos2, nossos3, nossos4}.
Calculating the perturbed vacuum field equations,
the following equation for the 4-dimensional components is obtained:
\begin{eqnarray}
&\partial_\nu\partial_\mu {h^\rho}_\rho - \partial_{(\nu}\partial^\rho h_{\mu)\rho} +\Box h_{\mu\nu} - \eta_{\mu\nu} \zeta^{\prime}e^{2\zeta}{{h^\rho}_\rho}^{\prime}   \nonumber
\\
&-e^{2\zeta}[h_{\mu\nu}^{\prime\prime} + 4\zeta^{\prime}h_{\mu\nu}^{\prime} + 2(4{\zeta^{\prime}}^2 + \zeta^{\prime\prime})h_{\mu\nu}] = 0.
\end{eqnarray}
Using further gauge conditions $\partial_\mu{h^\mu}_\nu =  0$, for each ${\footnotesize{\nu = 0,1,2,3}}$, and ${h^\mu}_\mu = 0$,
the equations above simplify to:
\begin{equation}\label{pert equation}
e^{2\zeta}\left[h_{\mu\nu}^{\prime\prime} + 4\zeta^{\prime}h_{\mu\nu}^{\prime} + 2(4{\zeta^{\prime}}^2 + \zeta^{\prime\prime})h_{\mu\nu}\right]-\Box h_{\mu\nu}= 0.
\end{equation}

If the perturbation $h_{\mu\nu}$ is written as the product
\begin{equation}\label{separation}
h_{\mu\nu}(y,x^\rho) = \phi(y)\chi_{\mu\nu}(x^\rho),
\end{equation}
and substituting Eq.(\ref{separation}) in Eq.(\ref{pert equation}), the usual 4-dimensional equations are obtained:
\begin{equation}\label{4D equation}
\Box \chi_{\mu\nu} + n^2 \chi_{\mu\nu} = 0,
\end{equation}
where $n$ is a constant of separation. The solution of this
equation is a linear superposition of plane waves:
\begin{equation}\label{4D solution}
\chi_{\mu\nu} = \epsilon_{\mu\nu}\exp(ik_\sigma x^\sigma),
\end{equation}
where the 4-wave vector satisfies $k_\mu k^\mu = n^2$.
This condition upheavals
 some informations with respect to the associated plane waves (\ref{4D solution}), for example the
dispersion relation associated. In this context, an extra dimension
dependence in the warp factor and consequently on the metric perturbation
affects the plane waves propagation modes.

The function $\phi(y)$ satisfies the
 equation
\begin{equation}\label{5D equation}
\phi^{\prime\prime} + 4 \zeta^{\prime}\phi^\prime + 2(4{\zeta^\prime}^2 + \zeta^{\prime\prime} + n^2e^{-2\zeta})\phi = 0,
\end{equation}
that describes the dependence of the metric perturbation
with respect to $y$, once specified a particular warp factor $\zeta(y)$.
It is interesting to note that the 4-dimensional solution
(\ref{4D solution}) is independent of the function $\zeta(y)$.

Now, in order to find a solution for $\phi(y)$ the function $\zeta(y)$ is substituted by the
solution (\ref{warp}) presented in \cite{RS,RS1}.
Equation (\ref{5D equation}) reads:
\begin{equation}\label{5d equation 2}
\phi^{\prime\prime} - 4k\phi^{\prime} + 2(n^2e^{2ky} + 4k^2)\phi = 0.
\end{equation}

Using the suitable change of variables $\xi=\frac{\sqrt{2}}{k}e^{ky}$, the
solution of the equation above can be written as:
\begin{equation}\label{5D solution}
\phi(\xi) = \xi^2\left[C_1 J_{2i}(n\xi) + C_{2}Y_{2i}(n\xi)\right],
\end{equation}
where the constants $C_1$ and $C_2$ can be suitably chosen in
order to reproduce the standard general relativity at the brane.
Indeed, as the $2^{\rm nd}$ order Bessel function $Y_{2i}(n\xi)$
diverges as $n \rightarrow 0$, in order to obtain physical solutions
we impose the constant $C_2 \equiv 0$.

Therefore, in the case where $n=0$ we have $k_\mu k^{\mu} = 0$ and the solution for $\phi$ is
\begin{equation}
\phi(y) = e^{2ky}\left(A_1e^{2iky} + A_2e^{-2iky}\right),
\end{equation}
where $A_1$ and $A_2$ are complex constants. Therefore, the well known case leading to general relativity is immediately obtained when
$h_{\mu\nu}(y=0,x^\rho) \equiv\chi_{\mu\nu}(x^\rho)$ in Eq.(\ref{separation}).
Indeed, when $n=0$ Eq.(\ref{4D equation}) implies such limit case.

The functions $J_{2i}$ and $Y_{2i}$ are Bessel functions of the
first and second kind of complex order $2i$, respectively.
The real part of (\ref{5D solution})
represents the dependence of the gravitational waves amplitudes
with the extra dimension $y$. Its profile is related to the
separation constant $n$ and to the constant $k$, which
characterizes the Anti-de Sitter length scale. The smaller the
$\phi(y)$, the bigger is $k$.

Furthermore, in spite of the behavior of
(\ref{5D solution}) to large values of $y$, from the Randall-Sundrum model compactification topology $S^1/\mathbb{Z}_2$ \cite{Horava1996, Horava19961, Polchinski1996}, the increasing of the
function $\phi(y)$ would be limited by the extra dimension
periodicity.
Note that the increasing behavior of $\phi(y)$ is also constrained
by the ratio between the separation constant and the
compactification radius of the extra dimension $n/k$.

Following the analysis in \cite{RS,RS1} and \cite{Horava1996, Horava19961}
 where the radius of compactification is of the
order of Planck scale, we can consider the solution
(\ref{5D solution}) in that regime. In this length scale the profile of solution in its asymptotic limit
$y\rightarrow0$ that presents a behavior given by expression in
function of Bessel function of first kind $J_\alpha(x)\sim\frac{1}{\Gamma(1+\alpha)}(\frac{x}{2})^\alpha$
with $\alpha\in\Im$. The same behavior happens when one analyze
the Bessel function of second kind $Y_\alpha$, that can be
written in terms of $J_\alpha$.

Standard classical 4-dimensional gravitational waves solution
is obtained in the case when we consider $y=0$, and we have in this case no extra dimension and the Minkowski metric.
This can be seen substituting $y=0$ in Eq.(\ref{5D equation}), where we obtain oscillatory solutions describing the gravitational waves.

\section{Concluding remarks and outlook}\label{concl}

This work analyzed gravitational waves in scenarios where our
universe is considered as a 3-brane embedded in five dimensions in accordance with Randall-Sundrum
model. Essentially, we analyzed the metric perturbations around an AdS$_5$ metric, considering the Randall-Sundrum solution for the
warp factor. The solution obtained is a linear combination of
Bessel functions of first and second kind with complex order.
Analyzing the real part of the solution, its behavior is
characterized by the ratio between the separation constant and the
constant $k$ which is of the order of the Planck length scale. The
solution is consistent in the case of small scale length of extra
dimension.

The prominent features of the analytical solutions in the Randall-Sundrum model: the extra dimension can be regulated by a separation
constant different from zero, which implies also on the propagation of
gravitational waves. At any rate the solution provides a
dependence, in fact, of extra dimension on waves propagation. The
propagation modes are affected by feature of topology of extra
dimension and its particular and main features can be represented
by real part of solution.
The solution obtained here can be also consistent if one considers
the ${S^1}/\mathbb{Z}_2$ topology on the extra dimension, which make
it periodic.

There is a question about the scenario in which are explored the
gravitational waves. This is related mainly to the separation
constant $n$. The whole spectrum of modes parameterized by it
would imply in a dispersion relation characteristic.

These modes can, in fact, characterize the gravitational waves
in scenarios with extra dimension.
The meaning of these modes could be interpreted like resonant
modes in black hole perturbation theory context for example.
Some additional aspects can be seen, e. g., in \cite{gog1,gog2}.
There the equations governing gravity wave propagation support
resonant solutions that are classified as {quasinormal
modes}. These resonant effects appear in many branches of physics
and can, as was treated in this context, be used to characterize
features of a given treatment metric perturbation.

Last but not least, recall that gravitational waves in general
relativity present two polarization states, namely, the ``+" and
the ``-". In the present case, however, since the wave equation
is different from that of the General Relativity, additional
polarization states can well exist.

Therefore, the existence of additional states of polarizations could
well give us some tips concerning the existence of extra dimensions.
In a paper to appear elsewhere we will consider such an issue.

\section*{Acknowledgments}
ACF thanks Prof. Oswaldo D. Miranda and Odylio D. Aguiar for
helpful discussions and INPE for the financial support and for the
hospitality, besides the post-doc grant of UFABC. MESA would like to thank the Brazilian Agency FAPESP
for support (grant 06/03158-0). JCNA would like to thank the
Brazilian agency CNPq for partial support (grant 307424/2007-3).
R. da Rocha is grateful to CNPq Projects 476580/2010-2 and
304862/2009-6 for financial support.

\end{document}